TITLE PAGE

**FULL TITLE:** An open-source treatment planning system for preclinical and clinical electron ultra-high dose rate radiotherapy

**RUNNING TITLE:** Treatment Planning System for Electron FLASH

**AUTHOR NAMES:**

Dixin Chen, MS[1,*]

Joseph B. Schulz, MS[1,2,*]

Angela Corvino, PhD[3]

Hamza Salman[1,4]

Zi Yang, PhD[1]

Stavros Melemenidis, PhD[1]

Lawrie Skinner, PhD[1]

Billy W. Loo Jr, MD, PhD[1]

M. Ramish Ashraf, PhD[1]

**CO-FIRST AUTHORS:**

Dixin Chen, MS

Joseph B. Schulz, MS

**SENIOR/CORRESPONDING AUTHORS:**

M. Ramish Ashraf, PhD[1]

Department of Radiation Oncology, Stanford University School of Medicine, Stanford, CA, USA, 94305

ramish.ashraf@stanford.edu

**AUTHOR FOR EDITORIAL CORRESPONDENCE:**

Dixin Chen, MS, dixin@stanford.edu

**INSTITUTIONS:**

1. Department of Radiation Oncology, Stanford University School of Medicine, Stanford, CA 94305, USA

2. Department of Medical Physics, University of Wisconsin-Madison, Madison, WI 53705, USA

3. Department of Surgery, Stanford University School of Medicine, Stanford, CA 94305, USA

4. Department of Physics and Astronomy, Vanderbilt University, Nashville, TN 37235, USA

**AUTHOR NOTES:**

*: Co-first authors.

**AUTHOR CONTRIBUTIONS:** All authors met the International Committee of Medical Journal Editors (ICMJE) criteria for authorship (https://www.icmje.org/recommendations/browse/roles-and-responsibilities/defining-the-role-of-authors-and-contributors.html). Dixin Chen and Joseph B. Schulz are co-first authors, each of whom is appropriately listed first when referencing this publication on their respective CVs.

**DATA AVAILABILITY:** The FLASH TPS source code is available at https://github.com/wizrads/FLASHTPS.

**Abstract**

**Purpose:** To develop an open-source treatment planning system (TPS) for electron ultra-high dose rate (UHDR) radiotherapy on a clinical linear accelerator and evaluate its planning-to-delivery workflow in mouse and anthropomorphic head phantoms.

**Methods and Materials:** FLASH TPS is a Python application that plans a treatment through CT import, contouring and target selection, beam's-eye-view collimator design, Monte Carlo (MC) dose calculation, reference-point normalization, and plan evaluation in a single workflow. The MC beam model represents a clinical linac in UHDR configuration with a reusable phase space scored below the treatment head. For beam model validation, simulated depth-dose curves and lateral dose profiles were compared with EBT-XD film measurements in solid water using local gamma analysis. Two end-to-end tests were then performed. An anatomically realistic 3D-printed mouse phantom was irradiated through a TPS-designed bilateral-lung collimator with film in a coronal plane through the lungs, and an anthropomorphic RANDO head phantom was irradiated through a nose-conformal collimator with bolus and film between axial sections at the nose.

**Results:** Local gamma passing rates for beam model validation in solid water pooled over seven depths were 97.6%, 99.4%, and 99.8% at 2%/2 mm, 3%/3 mm, and 4%/4 mm. Bilateral-lung plan passing rates were 95.5% at 1%/1 mm and 100% at 2%/2 mm. Nose plan passing rates were 98.5% at 2%/2mm and 100% at 3%/3 mm. Delivered doses measured by films were 21.9 Gy for the mouse plan (22 Gy prescribed to the planned target volume cross-section at the film plane) and 15.2 Gy for the head plan (15 Gy prescribed to the 90% isodose). Planned fraction of target volume irradiated above 40 Gy/s was 93.9% for the lung target and 99.9% for the nose target.

**Conclusions:** FLASH TPS workflow was validated end-to-end in mouse and anthropomorphic head phantoms. The TPS enables accurate electron FLASH planning on a clinical linac and provides an open-source platform for preclinical studies and future clinical translation.

## 1. INTRODUCTION

Ultra-high dose rate (UHDR, ≥40 Gy/s) irradiation can spare normal tissue relative to conventional dose rates while maintaining tumor control[1-3]. Since the first patient treatment in 2019[4], electron beams from dedicated accelerators and converted clinical linacs have remained the most accessible route to preclinical and early clinical FLASH studies[5-8].

Treatment planning, however, has not kept pace with delivery. Preclinical and veterinary UHDR electron treatments have so far relied on single standardized fields, without the treatment planning that is routine in conventional radiotherapy[9]. For instance, in one veterinary trial, late toxicity was linked to dose hot spots from tissue heterogeneities that a 3D dose calculation would have revealed[9,10].

Existing planning tools do not fit these machines. UHDR electron configurations, including those implemented on linacs in clinical use, differ from conventional electron beams in treatment head configuration, dose per pulse, and collimation. These differences limit the applicability of vendor beam data and commercial dose calculation algorithms developed for conventional delivery. Few commercial planning systems calculate or report dose-rate distributions, even though dose rate defines the FLASH irradiation condition. To work around these limitations, investigators build custom Monte Carlo (MC) beam models, manually edit simulation input files, and use separate scripts to calculate dose-volume metrics[9]. Additionally, for anatomically complex targets, one practical challenge is to turn a CT-based MC plan into a deliverable field by designing and fabricating a target-conformal collimator, then testing whether the delivered spatial dose pattern matches the plan.

An MC beam model of a converted clinical linac has been commissioned in a commercial planning system[11] and MC models have been reported for dedicated UHDR accelerators[12]. To our knowledge, however, no openly available tool that is independent of a commercial planning system integrates all these steps, from planning CT to a fabricated collimator and a verified delivery. This hinders the reproducibility of plans across institutions[13].

Here we present FLASH TPS, an open-source treatment planning system built to address this gap. It integrates the planning workflow from CT visualization and contouring through target-conformal collimator design, TOPAS MC dose calculation, reference-point normalization, and dose-volume and dose-rate-volume analysis. In addition, we describe the first MC beam model of the TrueBeam UHDR electron configuration[8], validated against film measurements in water-equivalent slabs. Finally, we demonstrate the complete workflow, from planning CT to film measurement of the delivered dose, at both small-animal and human scale in two end-to-end tests: 1) a bilateral-lung plan in an anatomically realistic mouse phantom and 2) a nose plan in an anthropomorphic RANDO head phantom.

## 2. METHODS AND MATERIALS

### 2.1 FLASH delivery platform

All irradiations used the reversible UHDR electron configuration of a clinical TrueBeam described previously[8]. Briefly, in service mode the 15 MV energy-switch setting is loaded, the photon target and monitor chamber are retracted, a clinically unused 4 MeV scattering foil is placed in the beam, and gun and radiofrequency settings are tuned to maximize dose per pulse. Restoring the standard settings returns the machine to routine clinical use. The resulting electron beam has a nominal energy of 12.6 MeV (determined from the measured $R_{50}$) and delivers up to 1.5 Gy per pulse at a pulse repetition frequency (PRF) of 180 Hz. Pulse charge is monitored with an in-line AC current transformer (ACCT) mounted on the linac head. For each collimator configuration, the ACCT charge reading is calibrated to dose with EBT-XD radiochromic film (Ashland, Wayne, NJ), and the gun voltage and the number of pulses is set to deliver the prescribed dose.

### 2.2 Monte Carlo beam model

The treatment head was modeled in TOPAS (version 4.0-4.2)[14,15], a Geant4-based MC toolkit, with the `g4em-standard_opt4` physics list. Simulation details are summarized in Table S1 following the AAPM TG-268 recommendations for reporting MC studies[16]. The primary electron beam is defined at the exit-window plane as a Gaussian pencil beam normally incident on the central axis. The components downstream of the electron source include, in beam order, a tantalum primary scattering foil, the clinically unused 4 MeV stepped aluminum secondary foil, the tungsten-alloy Y and X jaws, the copper ACCT toroid with its upstream polylactic acid (PLA) frame, and the collimator assembly (Fig. 1). The collimator assembly consists of a 7.1 mm PLA mounting layer, a 14 mm Cerrobend layer, and a 2 mm PLA holder layer.

The source uses a mean incident energy of 13.8 MeV with a 0.1% Gaussian energy spread at the exit window (Table S1). The treatment head was simulated once with $10^8$ primary histories, and a phase space was scored on a plane below the jaws and above the ACCT ($z = 57.4$ cm) containing 18.3 million particles. This phase space served as the source for all patient-specific simulations, which therefore model only the ACCT, the collimator assembly, and the patient or phantom.

### 2.3 Beam-model validation in water-equivalent slabs

The beam model was validated against EBT-XD film measurements in a solid-water phantom irradiated through a collimator with a $15 \times 15\ \mathrm{mm}^2$ square aperture with UHDR beam from the FLASH-configured TrueBeam (Fig. 2a). The phantom was centered on the beam axis with a 7 mm air gap between the collimator exit and the first film. Seven films were interleaved with six 2-mm solid-water slabs and backed by 100 mm of solid water, placing the films at nominal depths of 0, 2, 4, 6, 8, 10, and 12 mm (excluding film thickness). Films were scanned 24 h after irradiation on an Epson Expression 10000 XL flatbed scanner (Seiko Epson Corporation, Nagano, Japan) at 150 dpi (0.169 mm pixel spacing) without color correction. Film calibration and dose conversion followed the procedure described previously[8].

The phantom was modeled in TOPAS with `G4_WATER` for the solid-water slabs and each film as a 25 µm active layer between two 125 µm polyester substrates (0.275 mm total). Dose to medium was scored in the film active layers over a $20 \times 20\ \mathrm{mm}^2$ area at $0.2 \times 0.2\ \mathrm{mm}^2$ resolution and smoothed with a Gaussian kernel ($\sigma = 0.3$ mm). Crossline (*x*) and inline (*y*) profiles were extracted by averaging over a 1-mm-wide strip perpendicular to each profile, and depth dose was taken as the mean dose in the central $10 \times 10\ \mathrm{mm}^2$ of each film. A single normalization factor matched the simulated to the measured central mean dose at the shallowest film and was applied unchanged to all seven depths. To compare field shape independently of absolute dose, individual profiles were additionally normalized to their mean dose within the central $\pm 5$ mm. Agreement was evaluated with local gamma analysis using the film as the reference distribution, with the dose tolerance taken relative to the measured dose at each point and points below 10% of the 99th-percentile film dose excluded. Two-dimensional gamma analysis was performed at 2%/2 mm, 3%/3 mm, and 4%/4 mm for each film plane, and one-dimensional gamma analysis at 1%/1 mm and 2%/1 mm along the profiles.

## 2.4 FLASH TPS

### *2.4.1 Planning workflow*

FLASH TPS combines CT and structure visualization, contouring, collimator design, Monte Carlo simulation, and dose evaluation in a single application. Planning begins on the *CT* tab, where the patient CT is loaded and structures are imported or contoured directly in axial, sagittal, and coronal views. A planning target volume (PTV) is then defined and used in the *Collimator* tab to design an aperture in the beam's eye view, together with the source-to-surface geometry and patient offsets. Sending the collimator to the *Monte Carlo* tab writes the TOPAS aperture file and patient offsets and launches the simulation, which returns the dose grid together with its voxel-wise statistical uncertainty. The dose is passed back to the *CT* tab for normalization to the prescription and visual review as a color wash, isodose lines, and line profiles, and to the *DVH* tab, where pulse timing converts dose to dose rate and dose-volume histogram (DVH) and dose-rate volume histograms (DRVH) curves are computed for each structure. Each step is tracked on the *Home* tab, and the completed plan is saved as a study file that can be reopened for further review.

### *2.4.2 Reference-point dose normalization*

Raw MC output is expressed in relative dose units and scaled to the prescription dose $D_{\mathrm{Rx}}$ with one of three anchors, namely the mean dose inside a sphere (default radius 2.0 mm) around a user-selected point, the mean dose in the same sphere centered on a selected structure centroid (default mode), or a structure DVH statistic (from $D_5$ to $D_{99}$, or mean, default $D_{95}$). The scale factor is

$$s = \frac{D_{\mathrm{Rx}} f_{\mathrm{target}}}{D_{\mathrm{anchor}}}, \qquad (1)$$

where $f_{\text{target}}$ is the fraction of the prescription dose assigned to the anchor and $D_{\text{anchor}}$ is the raw MC dose at that anchor. Because the scale factor is an exact constant, the dose grid and any associated grid of absolute statistical standard errors are multiplied by the same factor, so relative uncertainties are unchanged.

The collimator-specific calibration establishes the absolute dose, whereas the Monte Carlo model provides only the relative dose distribution, which the TPS scales to the prescription with Eq. (1). Applying prescription normalization to an uncalibrated MC distribution therefore yields a planned relative dose scale rather than an absolute one. Finally, for the gamma analysis, the film and MC distributions are normalized independently so that their mean-dose difference is removed, and only spatial agreement is evaluated.

#### *2.4.3 Collimator design*

The *Collimator* tab derives a conformal collimator for an anterior beam from the target structure selected in the CT tab. The target boundary is extracted on the coronal slice of largest cross-sectional area perpendicular to the beam axis, with all closed loops retained and centered on the central axis at the target centroid. The outline is expanded radially by a specified margin (default 0 mm), optionally replaced by its 2D convex hull (enabled by default to enclose multi-loop targets, or unchecked to preserve concavities such as a mediastinal notch), and optionally simplified with the Ramer-Douglas-Peucker[17] or Visvalingam-Whyatt algorithm. The collimator polygon renders live over a cached digitally reconstructed radiograph alongside the unsimplified target loops. Export writes a TOPAS `TsAperture` coordinate file (`.ap`) in millimeters about the central axis. Selecting "Send to Monte Carlo" writes the. ap file, transfers the lateral `Patient/TransX` and `Patient/TransY` offsets that align the target centroid to the central axis, and opens the *Monte Carlo* tab. Physical collimators were fabricated from the same polygon with *CollimatorCreator*, a companion open-source CAD generator (`build123d`/OCCT) that outputs printable holders and Cerrobend casting molds[18].

#### *2.4.4 Monte Carlo integration*

The Monte Carlo tab launches TOPAS using parameter files or a phase-space template and provides access to collimator and patient-position parameters. A geometry preview supports inspection of the beam, collimator, and phantom relationship. Production dose is subsequently loaded for evaluation. During the simulation, CT numbers are converted to tissue materials and mass densities using the stoichiometric calibration method in Schneider *et al*[19]. Simulation versions, source settings, scoring geometries, materials, histories, and available uncertainty information for the reported runs are summarized in Table S1.

#### *2.4.5 Dose-volume and dose-rate-volume analysis*

For dose-volume analysis, binary structure masks were mapped to the dose grid and cumulative DVHs were calculated from the masked voxel doses. Because the FLASH effect depends on delivery timing, the DVH tab also computes live DRVHs. From the specified PRF $f$, pulse width τ, and pulse count $N_p$ (defaults 180 Hz, 3.2 μs, and 1), total irradiation time is

$$T_{\text{total}} = \frac{N_p - 1}{f} + \tau, \qquad (2)$$

which yields voxel mean dose rate $\dot{D}_i = D_i / T_{\text{total}}$, structure dose per pulse $D_{\text{mean}}/N_p$, and intrapulse dose rate $D_{\text{mean}}/(N_p \tau)$. For a specified UHDR threshold $\dot{D}_{\text{th}}$ (default 40 Gy/s), FLASH volume coverage is

$$V_{\text{UHDR}} = \frac{100}{N} \sum_{i=1}^{N} \left[ \dot{D}_i \geq \dot{D}_{\text{th}} \right], \qquad (3)$$

the percentage of structure voxels above threshold. Time is defined from the start of the first pulse to the end of the last pulse, including interpulse intervals.

### **2.5 End-to-end test 1: Bilateral-lung irradiation in mouse phantom**

An anatomically realistic 3D-printed mouse phantom with soft-tissue-, bone-, and lung-equivalent materials[13] was scanned on a clinical CT scanner (Siemens Biograph mCT, 120 kVp, 0.129 × 0.129 × 0.6 mm voxels, 222 slices) and loaded into the TPS. The whole lungs were contoured and imported as the PTV. The trachea, transferred from a separate segmentation of the phantom by rigid registration of the lungs, was the organ at risk. A bilateral-lung collimator conforming to both lungs with a 1 mm margin while shielding the mediastinum was generated in the *Collimator* tab. The prescription for the reported analysis was 22 Gy to the PTV cross-section at the film plane. The mold and holder were produced with *CollimatorCreator* and printed in polyethylene terephthalate glycol (PETG) and PLA on a Bambu Lab X1 Carbon (Bambu Lab, Shenzhen, China), and Cerrobend was poured at 75 °C and cooled for 24 hours. The simulation used the phase-space source and the collimator assembly of Sec. 2.2 with the lung-conformal collimator.

The phantom was irradiated with UHDR beam through the cast collimator with a 10 mm air gap between the collimator exit and the phantom. An EBT-XD film was inserted in a coronal plane through the lungs, approximately 8 mm below the surface. The film was scanned and converted to dose as in Sec. 2.3. The TOPAS dose, scored on the CT grid, was averaged over 0.52 mm along the beam at the film depth, smoothed with a Gaussian kernel ($\sigma$ = 0.3 mm), and interpolated to the registered film grid.

For the simulation and measurement comparison, point-by-point agreement was evaluated as the signed dose difference (TOPAS minus film). Profiles along $x$ and $y$ were averaged over 1-mm-wide strips. Two-dimensional local gamma analysis was performed with the film as reference and TOPAS as the evaluated distribution at 1%/1 mm and 2%/2 mm, excluding pixels below 10% of the 99th-percentile film dose, and one-dimensional local gamma analysis at 1%/1 mm and 2%/2 mm was performed along each profile.

For plan evaluation, the TOPAS dose on the CT grid was scaled so that the mean dose in the PTV cross-section at the film plane equaled the prescription, and DVHs and DRVHs of the lung PTV and the trachea were computed as in Sec. 2.4.5. Dose-rate analysis assumed 1 Gy per pulse at the prescription reference and 180 Hz, giving 22 pulses and 116.7 ms with a 3.2 μs pulse width.

**2.6 End-to-end test 2: Nose irradiation in anthropomorphic head phantom**

An anthropomorphic RANDO head phantom, supported on a headrest for immobilization, was scanned with a bolus on the face on a GE Discovery ST (120 kVp, $0.977 \times 0.977 \times 1$ mm voxels, 317 slices) and loaded into the TPS. The bolus consisted of two 1 cm sheets, which thinned to approximately 1.2 cm where they stretched over the nose, as measured on the CT acquired with the bolus in place. Three BBs were attached on the phantom for setup reproducibility. A nose PTV (8.4 $cm^3$) and a 2 mm skin shell (31.7 $cm^3$) were contoured in 3D Slicer[20] and imported as NIfTI labels. A nose-conformal collimator with a 1 mm margin (13 vertices, $34 \times 43$ mm) was generated, and its Cerrobend mold and holders were printed and cast as above. The prescription was 15 Gy to the 90% isodose line, with 100% defined as the maximum dose on the central axis.

The phantom was irradiated with the same headrest and bolus as at CT simulation and a 5 mm air gap between the collimator exit and the bolus surface (Fig. 2c). For each of three deliveries, an EBT-XD film cut to the anterior contour of the head was placed between axial sections 4 and 5 at the level of the nose. The first film was used for the two-dimensional comparison. It was registered to the TOPAS dose scored in the film plane by rigid translation and rotation, and the TOPAS and film dose maps were smoothed with Gaussian kernels of $\sigma = 1.4$ mm and 1.3 mm, respectively. Because the films were cut by hand to the anterior contour of the head, the dose within a few millimeters of the cut edge is unreliable as the cut disturbs the active layer. The analysis region was therefore defined as the film area eroded by 3-4 mm from the cut edge, and all two-dimensional comparisons were restricted to this region. The PTV was contoured on the CT to the skin surface, thus it extends beyond the displayed dose boundary by about the erosion margin at the anterior surface of the nose. The PTV and skin shell were mapped onto the simulation CT by rigid registration of the two CT scans. For the film comparison, film and TOPAS doses were each normalized to their own mean dose in the PTV cross-section in the film plane.

Two-dimensional local gamma analysis was performed with the film as reference at 2%/2 mm, 3%/3 mm, and 4%/4 mm, excluding pixels below 10% of the 99th-percentile film dose. Lateral profiles at depths of 25, 35, and 50 mm below the bolus surface, averaged over 2 mm in depth, and the central-axis depth-dose curve, taken as the mean of the three films over a 4.6-mm-wide strip after scaling the second and third films to the first over the PTV depth range, were compared with TOPAS by one-dimensional local gamma analysis at 2%/2 mm and 3%/3 mm.

For plan evaluation, the three-dimensional TOPAS dose was scaled so that the 90% isodose on the central axis equaled the prescription, and DVHs and DRVHs of the PTV and skin shell were computed as in Sec. 2.4.5. Dose-rate analysis assumed 15 pulses at 180 Hz with a 3.2 µs pulse width, yielding 77.8 ms.

## 3. RESULTS

### 3.1 Beam-model validation

The TOPAS beam model reproduced the measured lateral profiles and depth dose for FLASH irradiation (Fig. 3). At 6 mm depth, the crossline (*x*) and inline (*y*) profiles agreed in field width, central plateau, and penumbra, with one-dimensional local gamma passing rates of 88.9% and 91.2% at 1%/1 mm and 100% for both at 2%/1 mm (Fig. 3a,b). The measured and simulated depth-dose curves, each normalized to its maximum, differed by 2.3 percentage points on average and by at most 4.4 percentage points at 8 mm depth (Fig. 3c). Two-dimensional local gamma passing rates pooled over all seven depths were 97.6%, 99.4%, and 99.8% at 2%/2 mm, 3%/3 mm, and 4%/4 mm. The lowest per-film passing rates were at 4 mm depth, 88.2% at 2%/2 mm and 95.9% at 3%/3 mm, and every film plane exceeded 98.8% at 4%/4 mm. Profiles and gamma maps at all seven depths are given in Supplementary Figs. S1 and S2.

### 3.2 End-to-end test 1: Mouse Phantom

The film and TOPAS dose distributions showed the same field shape and dose pattern (Fig. 4a,b). The mean film dose in the PTV cross-section was 21.9 Gy, within 0.5% of the 22 Gy prescription. The largest residual differences were in the high-gradient regions at the field edges and in the lung next to the shielded mediastinum, where the film dose exceeded the simulated dose (Fig. 4c). Two-dimensional local gamma passing rates were 95.5% and 100% at 1%/1 mm and 2%/2 mm (Fig. 4d). Along the *x* profiles at $y = 2.5$ and $-2$ mm, one-dimensional local gamma passing rates were 90.9% and 98.1% at 1%/1 mm and 100% at 2%/2 mm for both (Fig. 4e,f). Along the *y* profiles through each lung at $x = -5$ and 5 mm, one-dimensional local gamma passing rates were 89.5% and 100% at 1%/1 mm and 100% at 2%/2 mm for both (Fig. 4g,h). Field dimensions and relative dose distribution therefore agreed closely, with the largest differences next to the shielded mediastinum.

### 3.3 End-to-end test 2: Anthropomorphic Head Phantom

The film and TOPAS dose distributions in the axial plane through the nose are in good agreement (Fig. 5a,b). The PTV outline extends beyond the displayed dose at the nose tip because the outer margin of the hand-cut film was excluded from the analysis (Sec. 2.6) and the dose there was not assessed. The mean film dose in the PTV cross-section was 15.2 Gy for the first film, 2.0% above the planned 14.9 Gy in the same cross-section for the 15 Gy prescription to the 90% isodose. Because the prescription was set on the central axis, this comparison tests the ACCT-based delivery calibration together with the Monte Carlo mapping from the prescription point to the film plane. The dose difference, TOPAS minus film, averaged 0.4% of the PTV mean dose with a standard deviation of 2.6% over the assessed region (Fig. 5c). Two-dimensional local gamma passing rates were 98.5%, 100%, and 100% at 2%/2 mm, 3%/3 mm, and 4%/4 mm (Fig. 5d). For the lateral profiles at 25, 35, and 50 mm below the bolus surface, one-dimensional local gamma passing rates were 96.7%, 93.0%, and 74.1% at 2%/2 mm and 100%, 100%, and 88.3% at 3%/3 mm (Fig. 5e-g). The central-axis depth-dose curves of the three films agreed with one another to within 0.8% of the PTV mean dose, and their mean agreed with TOPAS at all assessed depths at both criteria, with $R_{80}$ of 38.5 and 39.1 mm and $R_{50}$ of 50.9 and 51.8 mm below the bolus surface for film and TOPAS, respectively (Fig. 5h).

### 3.4 Planned dose and dose rate

Figure 6 shows the planned dose distributions and the DVHs and DRVHs of the two plans. For the bilateral lung plan, normalized to a 22 Gy mean dose in the PTV cross-section at the film plane, the mean dose over the whole lung PTV was 19.3 Gy and the mean trachea dose was 1.9 Gy (Fig. 6a,c). For the 22-pulse delivery, the 40 Gy/s threshold corresponds to 4.7 Gy, and $V_{\mathrm{UHDR}}$ was 93.9% for the PTV and 12.8% for the trachea.

For the nose plan, along the central axis, the 90% isodose lay 27.8 mm below the bolus proximal surface and the distal edge of the PTV at 30.8 mm, so the deepest 3 mm of the PTV on the central axis received less than 90% of the prescription (Fig. 6b). The PTV received a mean dose of 14.2 Gy with a $D_{95}$ of 9.3 Gy, and the skin shell a mean dose of 3.0 Gy (Fig. 6d). For the 15-pulse delivery, the 40 Gy/s threshold corresponds to 3.1 Gy, and $V_{\mathrm{UHDR}}$ was 99.9% for the PTV and 25.0% for the skin shell.

## 4. DISCUSSION

FLASH TPS closes the gap between a UHDR-capable linac and a usable planning workflow. The entire workflow from CT import to a cast collimator and a verified dose distribution is completed in one application. Several assumptions are built into the beam model and define the configuration to which the phase space and the validation in Sec. 2.3 apply. First, both scattering foils are in the beam, as they were for every experiment reported here. The secondary foil could instead be replaced by an open carousel slot, which would increase the dose per pulse substantially at the expense of field flatness, but that configuration was not modeled. Second, the ACCT toroid used for pulse-charge monitoring is part of the beamline and is the limiting aperture downstream of the jaws, restricting the usable open field at the collimator plane to about 5.8 cm in diameter. Other charge-monitoring solutions would remove this constraint but would require a new downstream geometry and re-validation. Third, the jaws were fixed at a nominal $8 \times 8$ cm$^2$ field at isocenter for all simulations and experiments. Since the jaw opening is embedded in the phase space, any other setting requires the upstream simulation to be repeated. Fourth, components that are retracted or of negligible dosimetric consequence in the UHDR configuration (exit window, primary collimator, retracted target and monitor chamber, multileaf collimator, and field-light mirror) were omitted, and the model is perfectly coaxial. Finally, the model has no explicit pulse structure. Dose is scored per primary history, and dose rate is derived afterward from the pulse schedule (Sec. 2.4.5). The model also has no absolute output calibration. Absolute dose is set by the film-calibrated ACCT reading for each collimator, so a new collimator requires a new calibration measurement, while the TPS can plan any collimator in relative terms.

Several limitations constrain the interpretation of the results presented in this work. Because the film and MC distributions were normalized independently to their planar means, any difference in absolute dose between measurement and simulation was removed by construction, and the gamma pass rates therefore characterize spatial agreement alone rather than the accuracy of the predicted dose. The gamma results are further conditioned by the processing chain. Film smoothing reduces noise but blurs the penumbral gradients that dominate small-field agreement. Registration uncertainty of even a fraction of a millimeter shifts the comparison in these same high-gradient regions. On the simulation side, the dose is scored in voxels of finite size, which averages over the voxel and lowers the apparent gradient relative to film, while statistical noise in low-dose voxels is not separated from systematic disagreement by the gamma metric. The most important of these limitations, the absence of an absolute dose comparison, could be addressed with a modest extension of the existing workflow. A single reference measurement per collimator with a calibrated small-volume ion chamber or traceably calibrated film at a fixed depth in solid water, combined with the MC dose scored in the same 2.0 mm sphere already used as the normalization anchor, would yield a dose-per-history calibration factor. Substituting this measurement-derived factor for the prescription-derived scale factor in Eq. (1) would convert any MC distribution to absolute dose, and a second measurement at an independent depth or off-axis position would then allow absolute agreement to be reported alongside the spatial agreement presented here, while also quantifying any residual output difference that the planar-mean normalization currently conceals.

## 5. CONCLUSIONS

We have developed and released FLASH TPS, an open-source treatment planning system that integrates CT visualization and contouring, target-conformal collimator design, MC dose calculation, reference-point dose normalization, and dose-volume and dose-rate-volume analysis for electron FLASH on a FLASH-enabled clinical linac. The MC beam model reproduced measured depth dose and lateral profiles in solid water (97.6% local gamma passing at 2%/2 mm), and the full chain was verified end-to-end in an anatomically realistic mouse phantom (95.5% local gamma passing at 1%/1 mm) and an anthropomorphic head phantom (98.5% local gamma passing at 2%/2 mm). This work provides a shared platform for reproducible preclinical UHDR studies and a starting point for translational planning.

## FIGURE CAPTIONS

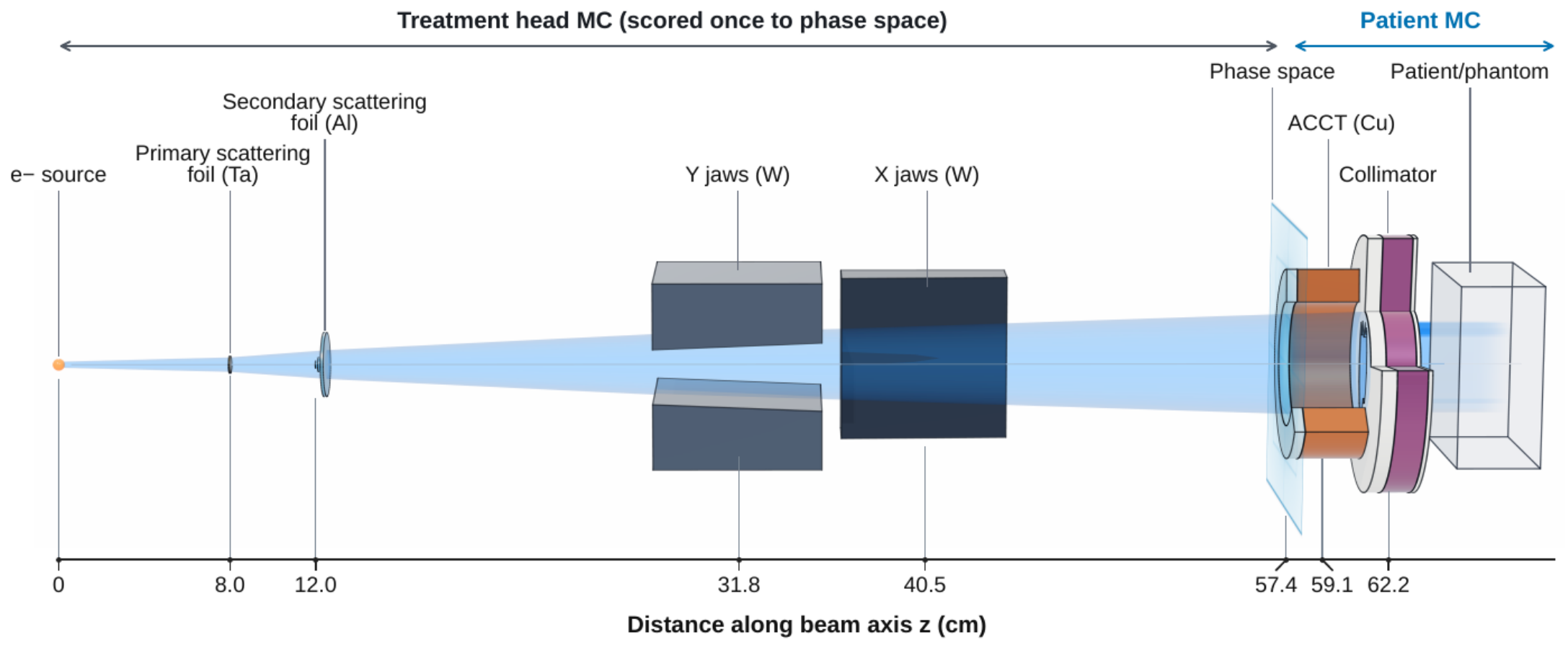

**Figure 1: Monte Carlo electron FLASH beam model.** Scale schematic of the modeled TrueBeam beamline in the UHDR electron configuration, with *z* measured along the beam axis from the electron source. The treatment head (Ta primary scattering foil, stepped Al secondary foil, and focused tungsten-alloy Y and X jaws) is simulated once and scored to a phase-space plane at $z = 57.4$ cm. Each patient-specific run then transports particles from the phase space through the AC current transformer (ACCT), modeled as a copper toroid (29 mm thickness) with a 5 mm PLA frame upstream, the collimator (7.1 mm PLA, 14 mm Cerrobend, 2 mm PLA) and the patient or phantom. Blue, electron beam envelope.

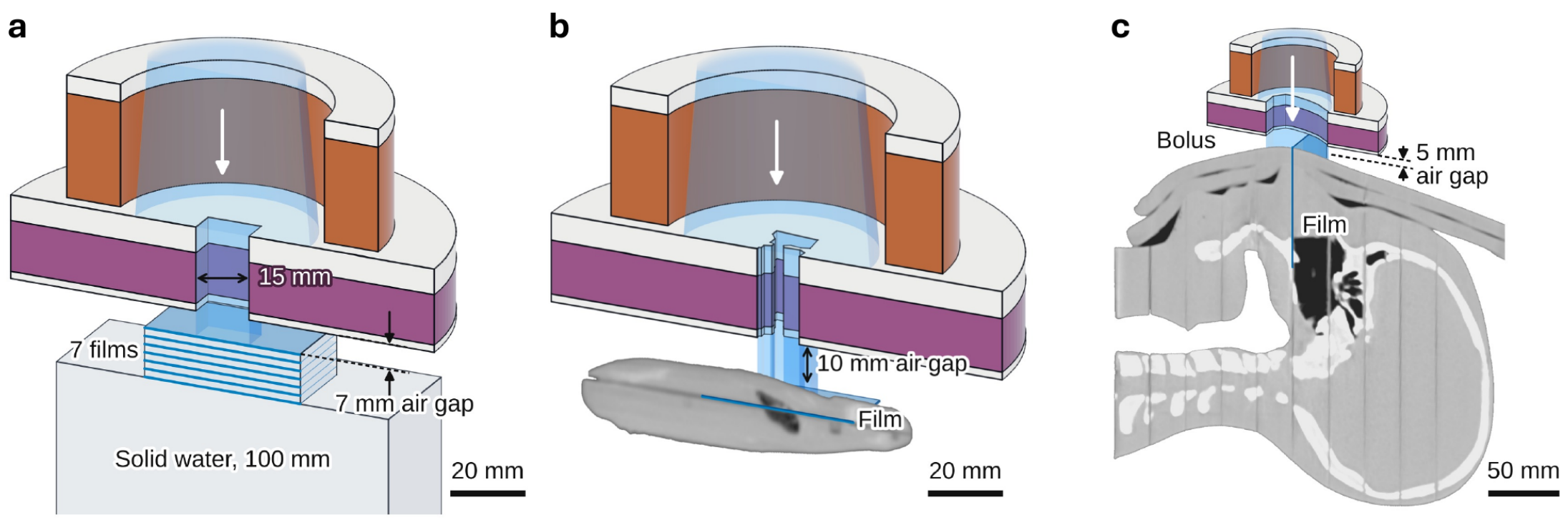

**Figure 2: Validation setups.** Cutaway views of the three setups, with the beam (white arrow) passing through the ACCT and the collimator. **a**, Beam-model validation: a film-solid-water stack irradiated through a $15 \times 15$ mm square aperture with a 7 mm air gap. Seven EBT-XD films are interleaved with 2 mm solid-water slabs to sample increasing depth, backed by 100 mm of solid water. **b**, Preclinical end-to-end test: a 3D-printed mouse phantom irradiated through the bilateral-lung aperture with a 10 mm air gap and a coronal film through the lungs. **c**, Clinical end-to-end test: a RANDO head phantom irradiated through the nose-conformal aperture with bolus and a 5 mm air gap, and a film between axial sections through the nose. Scale bars, 20 mm (**a**,**b**) and 50 mm (**c**).

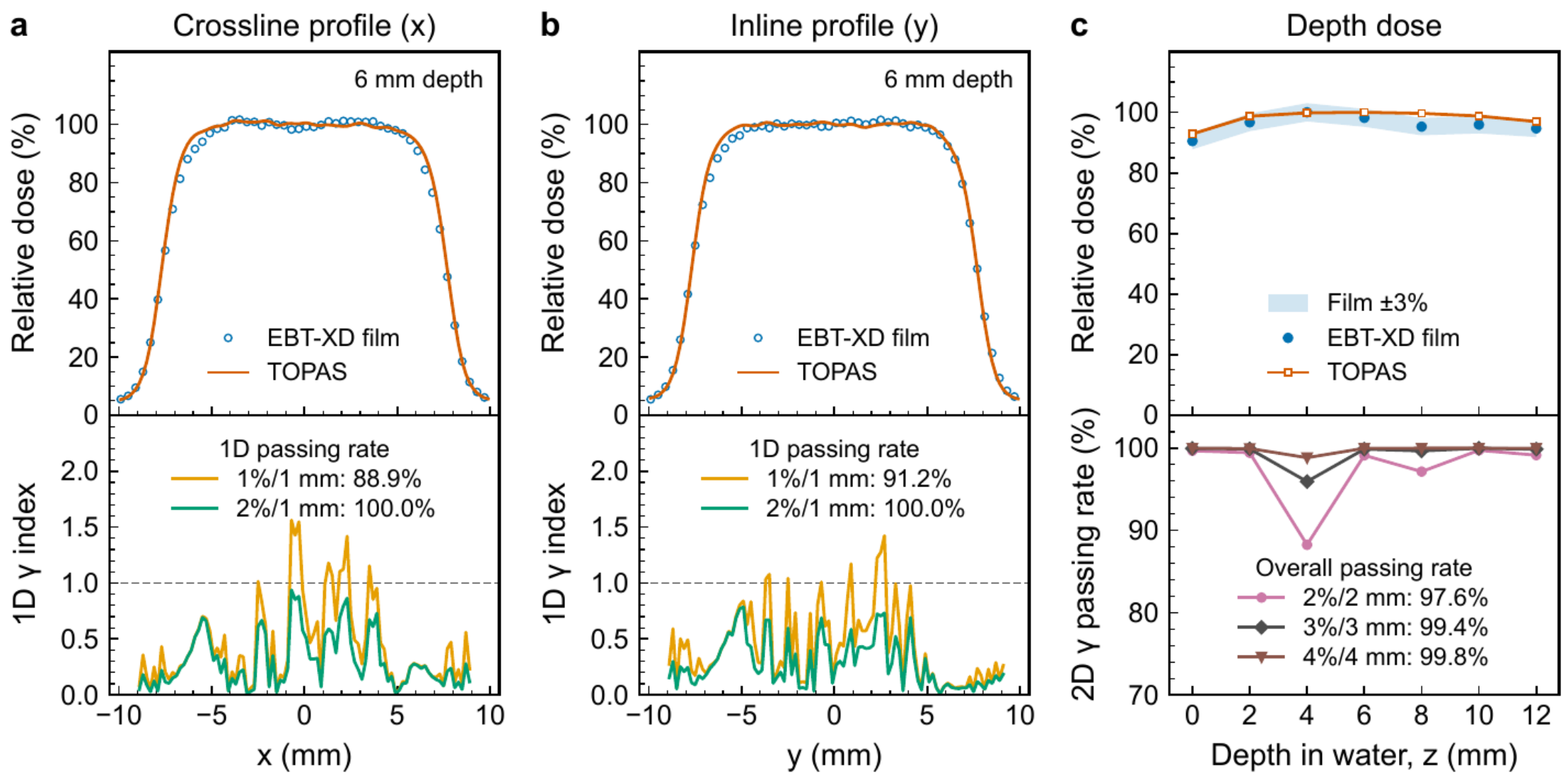

**Figure 3: Validation of the TOPAS electron beam model against film dosimetry in solid water. a**, **b**, Crossline (*x*) and inline (*y*) relative dose profiles at 6 mm depth for FLASH irradiation, measured with EBT-XD film and simulated with TOPAS (top), and the corresponding one-dimensional local gamma index at 1%/1 mm and 2%/1 mm (bottom). Dashed line, $\gamma = 1$. **c**, Depth dose from the mean dose in the central $10 \times 10$ mm$^2$ of each film, normalized to its maximum (top), and two-dimensional local gamma passing rate ($\gamma \leq 1$) for each film at 2%/2 mm, 3%/3 mm and 4%/4 mm (bottom). Shading, ±3% of the film dose. Points below 10% of each film's 99th-percentile dose are excluded from all gamma analyses. Legend values are passing rates along each profile and pooled over all seven depths.

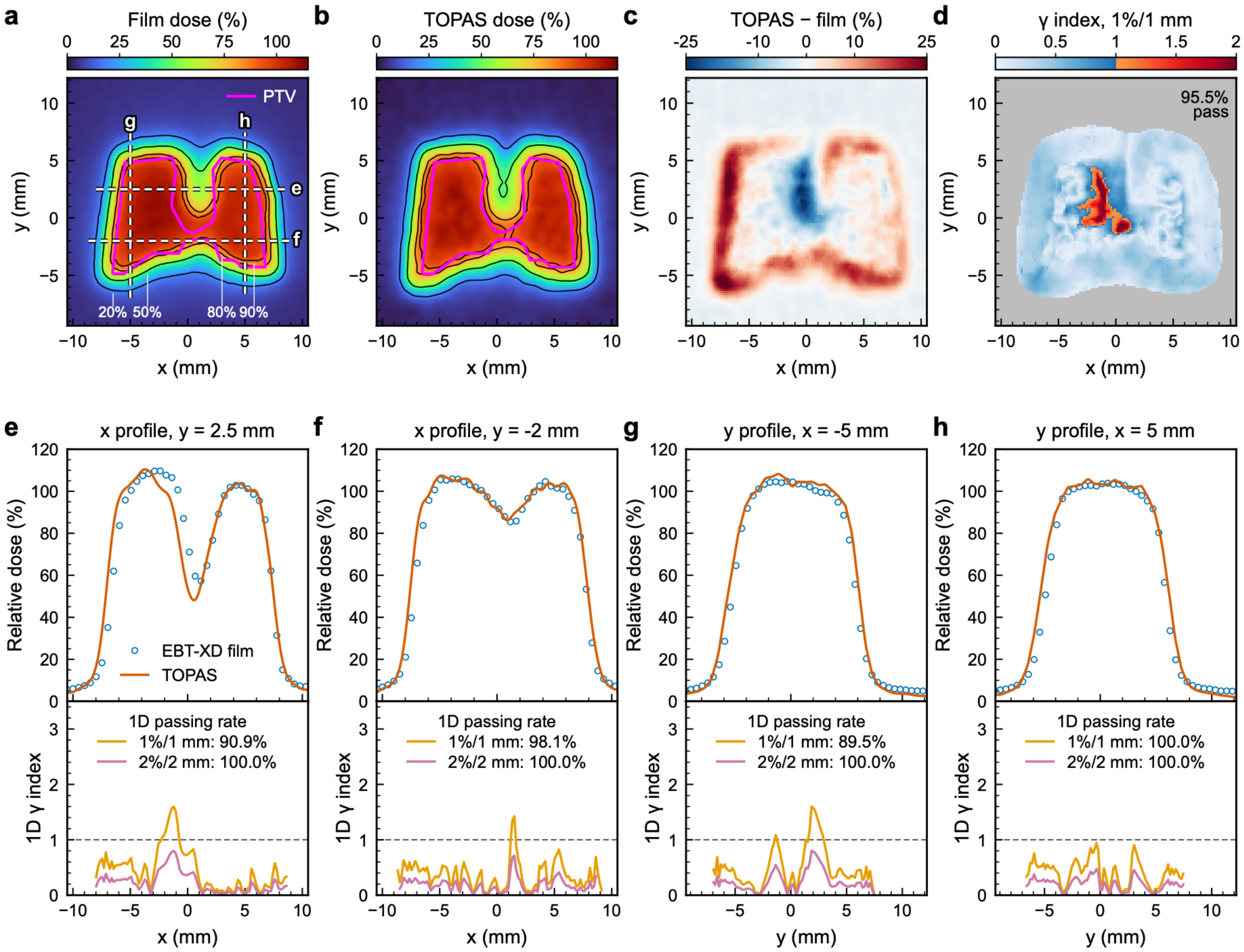

**Figure 4: End-to-end validation of a bilateral-lung FLASH plan in a 3D-printed mouse phantom. a**,**b**, Relative dose in the coronal film plane (depth ≈ 8 mm) measured with EBT-XD film (**a**) and simulated with TOPAS (**b**). Dashed lines, positions of the profiles in **e**-**h**. **c**, Dose difference (TOPAS minus film). **d**, Two-dimensional local gamma index at 1%/1 mm with 95.5% of pixels passing ($\gamma \leq 1$). Grey, below the 10% low-dose threshold and not assessed. **e**-**h**, Dose profiles along *x* (**e**,**f**) and *y* (**g**,**h**) for film and TOPAS (top), with the corresponding one-dimensional local gamma index at 1%/1 mm and 2%/2 mm (bottom). Dashed line, $\gamma = 1$. Legend values are passing rates along each profile.

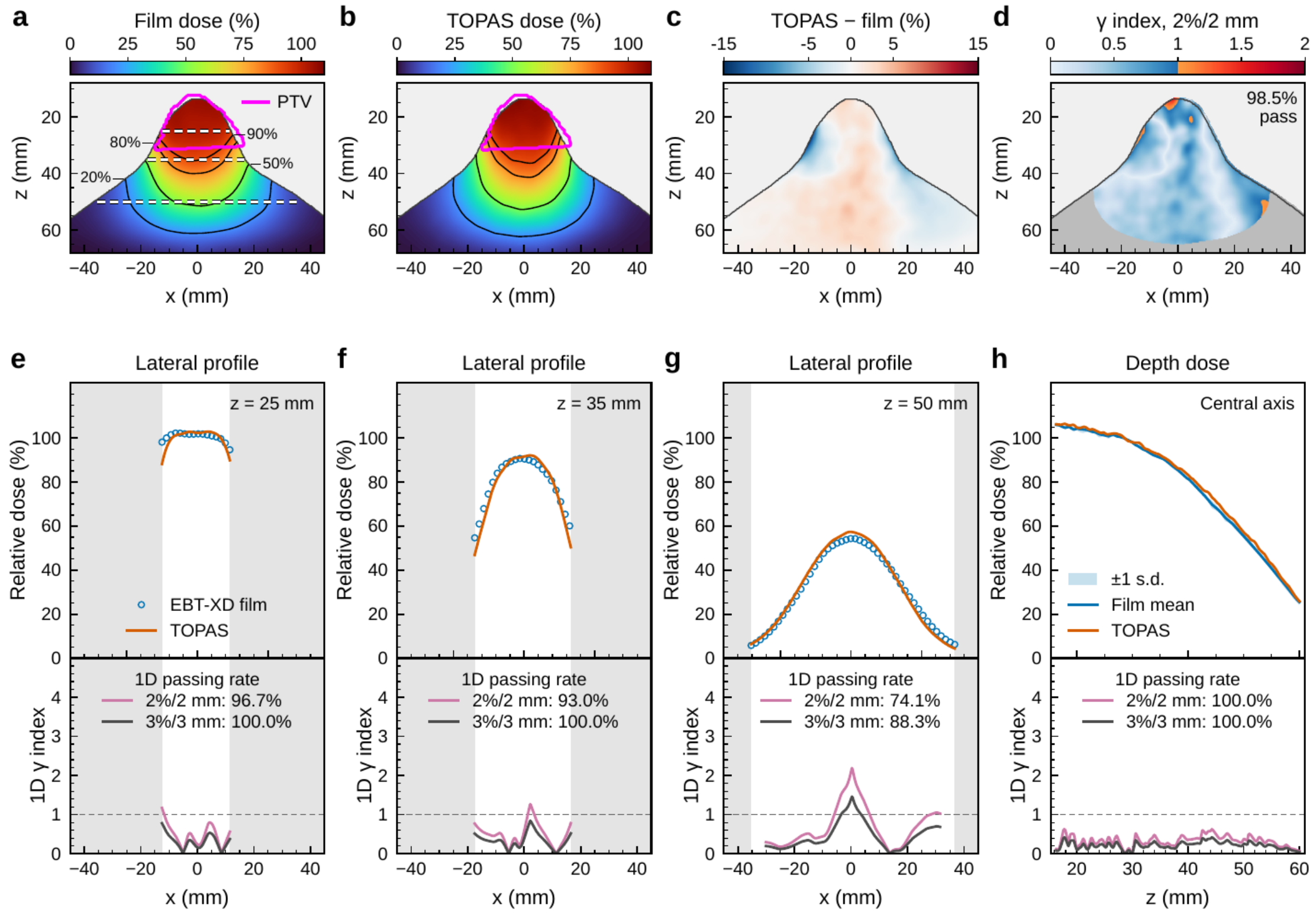

**Figure 5: End-to-end validation of a nose FLASH plan in an anthropomorphic head phantom. a**,**b**, Relative dose in the axial film plane measured with EBT-XD film (**a**) and simulated with TOPAS (**b**), normalized to the mean dose in the PTV cross-section. The dose is shown within the film analysis region, the hand-cut film eroded by 3-4 mm from its edge to exclude unreliable edge dose, so the PTV outline (magenta) extends beyond the displayed dose at the anterior surface of the nose. Dashed lines, depths of the profiles in **e**-**g**. **c**, Dose difference (TOPAS minus film). **d**, Two-dimensional local gamma index at 2%/2 mm with 98.5%, 100% and 100% of pixels passing ($\gamma \leq 1$) at 2%/2 mm, 3%/3 mm and 4%/4 mm, excluding regions with dose below 10% of the film's 99th-percentile dose. **e**-**g**, Lateral dose profiles at depths $z = 25$, 35 and 50 mm, averaged over 2 mm in $z$, for film and TOPAS (top), with the corresponding one-dimensional local gamma index at 2%/2 mm and 3%/3 mm (bottom). Grey shading, no film data. **h**, Central-axis depth dose for TOPAS and the mean of three films, averaged over a 4.6-mm-wide strip (top), with the one-dimensional local gamma index (bottom). Shading, ±1 standard deviation (s.d.) across films, which is smaller than the line width at most depths. Dashed line, $\gamma = 1$. Legend values are passing rates along each profile.

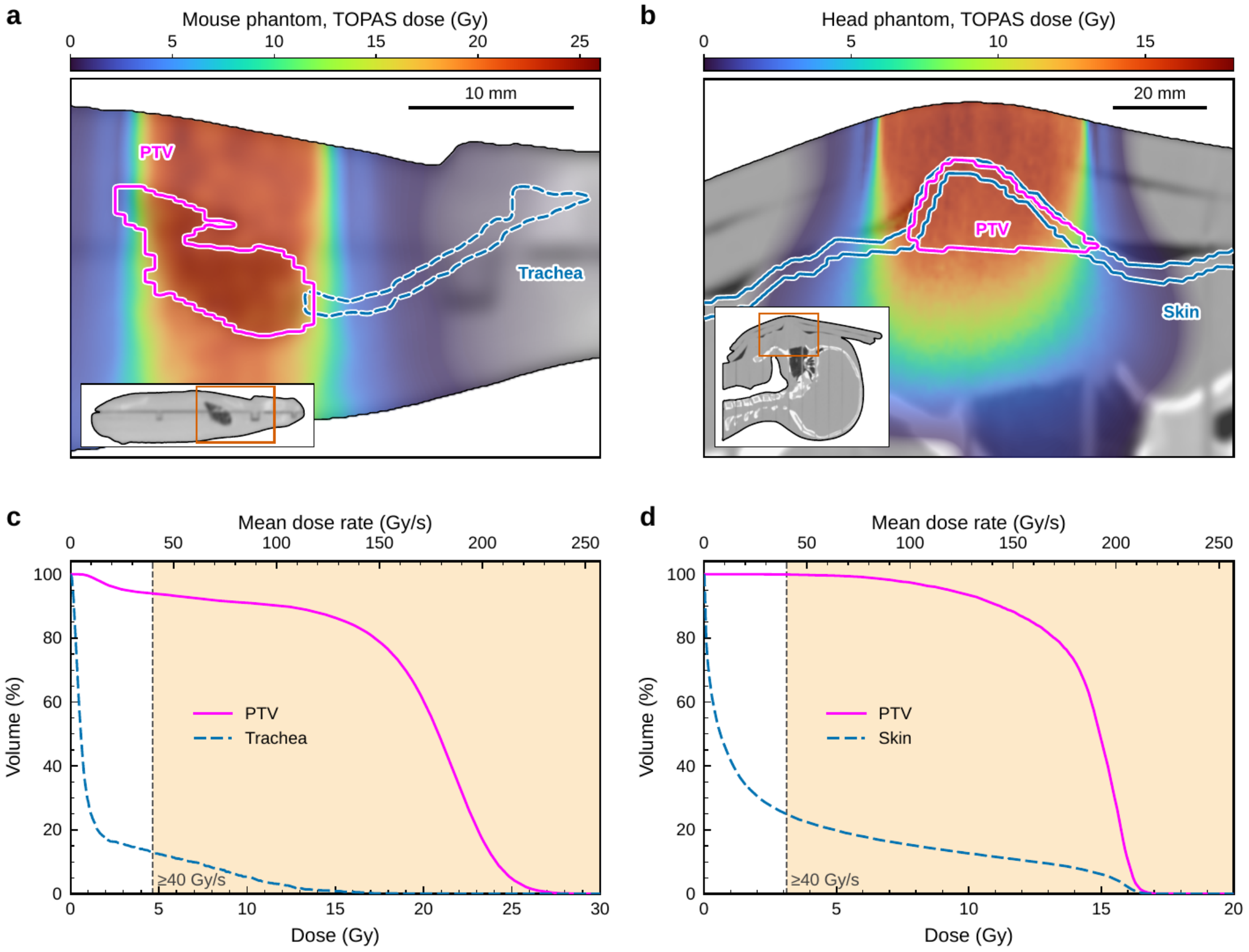

**Figure 6: Planned dose distribution and dose-volume histograms for the two end-to-end tests. a**,**b**, Sagittal CT slices of the mouse phantom (**a**) and the head phantom with bolus (**b**) with the TOPAS dose as a color wash and the outlines of the planning target volume (PTV, magenta) and the organ at risk (blue). The trachea in **a** lies on the midline and is projected onto the slice. Insets show the whole CT section, with the displayed region boxed. Scale bars, 10 mm (**a**) and 20 mm (**b**). **c**,**d**, Cumulative dose-volume histograms of the PTV and the organ at risk for the lung (**c**) and nose (**d**) plans. The top axis gives the mean dose rate, the dose divided by the delivery time, so each curve is also the dose-rate-volume histogram. Shading marks mean dose rates of at least 40 Gy/s assuming 1 Gy/pulse was delivered at 180 Hz.

## REFERENCES

1. V. Favaudon et al., Ultrahigh dose-rate FLASH irradiation increases the differential response between normal and tumor tissue in mice, Sci. Transl. Med. **6**, 245ra93 (2014).

2. P. Montay-Gruel et al., Irradiation in a flash: Unique sparing of memory in mice after whole brain irradiation with dose rates above 100 Gy/s, Radiother. Oncol. **124**, 365–369 (2017).

3. M.-C. Vozenin et al., The Advantage of FLASH Radiotherapy Confirmed in Mini-pig and Cat-cancer Patients, Clin. Cancer Res. **25**, 35–42 (2019).

4. J. Bourhis et al., Treatment of a first patient with FLASH-radiotherapy, Radiother. Oncol. **139**, 18–22 (2019).

5. E. Schüler et al., Experimental Platform for Ultra-high Dose Rate FLASH Irradiation of Small Animals Using a Clinical Linear Accelerator, Int. J. Radiat. Oncol. Biol. Phys. **97**, 195–203 (2017).

6. H. J. No et al., Clinical Linear Accelerator-Based Electron FLASH: Pathway for Practical Translation to FLASH Clinical Trials, Int. J. Radiat. Oncol. Biol. Phys. **117**, 482–492 (2023).

7. N. Esplen, M. S. Mendonca, and M. Bazalova-Carter, Physics and biology of ultrahigh dose-rate (FLASH) radiotherapy: a topical review, Phys. Med. Biol. **65**, 23TR03 (2020).

8. S. Melemenidis et al., Noninvasive Reversible Software-Based Electron FLASH Irradiation Configuration of a Linear Accelerator in Clinical Use, Int. J. Radiat. Oncol. Biol. Phys. **126**, 848–855 (2026).

9. M. Rahman et al., FLASH radiotherapy treatment planning and models for electron beams, Radiother. Oncol. **175**, 210–221 (2022).

10. C. Rohrer Bley et al., Dose- and Volume-Limiting Late Toxicity of FLASH Radiotherapy in Cats with Squamous Cell Carcinoma of the Nasal Planum and in Mini Pigs, Clin. Cancer Res. **28**, 3814–3823 (2022).

11. M. Rahman et al., Treatment Planning System for Electron FLASH Radiation Therapy: Open-Source for Clinical Implementation, Int. J. Radiat. Oncol. Biol. Phys. **112**, 1023–1032 (2022).

12. T. Dai et al., First Monte Carlo Beam Model for Ultra-high Dose Rate Radiotherapy with a Compact Electron Linac, Med. Phys. **51**, 5109–5118 (2024).

13. M. R. Ashraf et al., Multi-Institutional Audit of FLASH and Conventional Dosimetry With a 3D Printed Anatomically Realistic Mouse Phantom, Int. J. Radiat. Oncol. Biol. Phys. **120**, 287–300 (2024).

14. J. Perl, J. Shin, J. Schümann, B. Faddegon, and H. Paganetti, TOPAS: an innovative proton Monte Carlo platform for research and clinical applications, Med. Phys. **39**, 6818–6837 (2012).

15. B. Faddegon et al., The TOPAS tool for particle simulation, a Monte Carlo simulation tool for physics, biology and clinical research, Phys. Med. **72**, 114–121 (2020).

16. I. Sechopoulos et al., RECORDS: improved Reporting of montE CarlO RaDiation transport Studies: Report of the AAPM Research Committee Task Group 268, Med. Phys. **45**, e1–e5 (2018).

17. D. H. Douglas and T. K. Peucker, Algorithms for the Reduction of the Number of Points Required to Represent a Digitized Line or its Caricature, Cartographica **10**, 112–122 (1973).

18. J. B. Schulz et al., A software tool for semi-automated parametric collimator creation for linac-based preclinical irradiation studies, manuscript in preparation (2026).

19. W. Schneider, T. Bortfeld, and W. Schlegel, Correlation between CT numbers and tissue parameters needed for Monte Carlo simulations of clinical dose distributions, Phys. Med. Biol. **45**, 459–478 (2000).

20. A. Fedorov et al., 3D Slicer as an image computing platform for the Quantitative Imaging Network, Magn. Reson. Imaging **30**, 1323–1341 (2012).

# Supplementary Materials

## S1 MONTE CARLO REPORTING DETAILS

**Table S1:** Additional Monte Carlo simulation and computational details, following AAPM TG-268 reporting recommendations.

| Item | Description | | |
|---|---|---|---|
| Code and physics | TOPAS 4.0 (mouse and head runs) and TOPAS 4.2.p2 (solid-water run)<br>Electromagnetic physics g4em-standard_opt4 for all runs | | |
| Source characteristics | Electrons defined at the exit-window plane ($z = 0$), normally incident on the central axis. Gaussian energy distribution, mean 13.8 MeV, spread 0.1% ($1\sigma$). Circular Gaussian focal spot, $\sigma = 0.4$ mm in *x* and *y*, with a 100 mm elliptical position cutoff. Gaussian angular distribution, $\sigma = 40$ mrad in *x* and *y*, with a 55° angular cutoff. | | |
| Treatment-head run and phase-space recording | $1 \times 10^8$ primary histories, 20 threads. Single scoring plane at $z = 57.4$ cm (below the jaws, above the ACCT toroid), $50 \times 50$ cm², TOPAS ASCII format with time of flight, run, event, track and parent identifiers, creator process, vertex information and random-number seeds recorded for each particle. No particle-type or direction filters: all electrons, photons, and positrons crossing the plane were recorded. Jaws fixed at nominal $8 \times 8$ cm² at isocenter. Of the $10^8$ primaries, 16,476,222 produced at least one particle at the plane, giving 18,303,400 particle records (12,420,033 electrons, 5,871,894 photons, 11,473 positrons). Maximum kinetic energies 13.70 MeV (electrons), 13.80 MeV (photons) and 11.34 MeV (positrons). Empty histories were not replayed in the downstream runs (PhaseSpaceIncludeEmptyHistories = False). | | |
| Transport parameters | No production cuts, tracking cuts, step limits or physics regions were specified in any parameter file, so the TOPAS defaults applied: production range cut 0.05 mm for all particles in all volumes, electromagnetic tables from 100 eV to 500 MeV, and no maximum step size beyond the multiple-scattering step limitation of the chosen physics constructor. No variance-reduction techniques (no particle splitting, Russian roulette or range rejection). The only history reuse was the twentyfold replay of the phase space in the solid-water run. Geometry overlap checks were enabled for the solid-water and mouse runs and disabled for the head runs. | | |
| Patient and phantom geometry | CT voxels converted to tissue materials and mass densities with the with the Schneider conversion (ImagingToMaterialConverter = "Schneider"). Solid-water slabs and backing block as G4_WATER. Films as a 25 µm active layer (H 56.8%, Li 0.7%, C 27.6%, O 13.3%, Al 1.6%, 1.2 g/cm³) between two 125 µm polyester substrates (H 36.4%, C 45.5%, O 18.1%, 1.35 g/cm³). Collimator layers PLA ($C_3H_4O_2$, 1.24 g/cm³) and Cerrobend (Bi 50%, Pb 26.7%, Sn 13.3%, Cd 10%, 9.4 g/cm³), ACCT toroid G4_Cu. In the head runs, the ACCT and collimator were placed in a layered mass geometry parallel world so that they overlap the CT bounding box without geometry conflicts. | | |
| Computational resources | Single multithreaded TOPAS process per run, no cluster.<br>Solid-water run: macOS workstation, 8 threads.<br>Mouse run and head 3D dose run: macOS workstation, 10 threads.<br>Head film-plane run: laptop, 4 threads. | | |
| | **Final dose-calculation runs** | | |
| | **Solid water** | **Mouse phantom** | **Head phantom** |
| Phase-space histories used | All 16,476,222 non-empty histories, each replayed 20 times (329,524,440 transported histories) | All 16,476,222 non-empty histories, single pass | All 16,476,222 non-empty histories, single pass |

| Item | Description | | |
|---|---|---|---|
| Independent upstream histories represented | 100,000,000 upstream primaries; replay does not add independent upstream histories | 100,000,000 upstream primaries represented by the source file | 100,000,000 upstream primaries represented by the source file |
| Total transported phase-space particle records | 366,068,000 | 18,303,400 (per run) | 18,303,400 (per run) |
| Phase-space reuse factor | 20 | 1 | 1 |
| Random seeds / independent batches | TOPAS default seed, one batch | TOPAS default seed, one batch | TOPAS default seed, one batch per run |
| Scoring geometry | DoseToMedium in the seven film active layers, $100 \times 100$ bins of $0.2 \times 0.2$ mm² ($20 \times 20$ mm²), plus a parallel-world 0.2 mm grid over the stack ($32.4 \times 95.2 \times 28.8$ mm³) | DoseToMedium on the CT grid, $512 \times 368 \times 222$ voxels of $0.129 \times 0.129 \times 0.6$ mm³ | DoseToMedium on (i) a parallel-world film-plane grid of $500 \times 260$ bins (0.20 mm lateral × 0.35 mm in depth, 2 mm thick, $100 \times 90$ mm²) and (ii) the CT grid, $512 \times 512 \times 317$ voxels of $0.977 \times 0.977 \times 1.0$ mm³ |
| Reported voxel-noise estimates (%) | 2.8 per $0.2 \times 0.2$ mm² film voxel in the central $10 \times 10$ mm² plateau, before smoothing | 2.2 to 2.7 per CT voxel in the region above 80% of the maximum dose | 3.0 to 3.6 per film-plane bin and 2.6 to 4.1 per CT voxel in the region above 80% of the maximum dose |
| Scoring and uncertainty | All scorers report dose to medium summed over histories and are scaled to the prescription afterwards. For the solid-water run, the per-voxel TOPAS Standard_Deviation of the per-history dose was recorded, and the relative standard error of each voxel was taken as $\sqrt{N}\,\sigma_h/D$, with $N$ the number of transported histories. Because the 20 replays of one upstream history are correlated, this was cross-checked against the standard deviation of differences between adjacent voxels divided by $\sqrt{2}$, which gave the same value (2.8%). Mouse and head: no uncertainty scorer, so uncertainty was taken from adjacent-voxel differences in the region above 80% of maximum dose (2.2–2.7% and 2.6–4.1% per CT voxel, 3.0–3.6% per film-plane bin). These per-voxel values are upper bounds on the statistical uncertainty of the reported quantities, which smoothing and profile averaging reduce further. | | |

## S2 TREATMENT PLANNING SYSTEM SOFTWARE DETAILS

### S2.1 Architecture and workflow

FLASH TPS is written in Python (versions 3.10 to 3.12) with PySide6 for the graphical interface and VTK (≥9.5) for 3D image processing and rendering[1]. It relies solely on open-source Python packages and runs on macOS, Linux, and Windows. Five core tabs (*Home*, *CT*, *Collimator*, *Monte Carlo*, and *DVH*) share an in-memory workspace that holds the CT volume in canonical left posterior superior (LPS) orientation, the scaled dose and voxel wise $1\sigma$ statistical uncertainty grids, and binary structure masks registered to the CT. An auto-discovered plugin interface (`WorkspaceAPI`) mounts optional contributor tabs, such as automatic mouse CT segmentation on CPU and linear-quadratic radiobiology evaluation, without changes to the core code. On the *Home* tab, a six-step workflow strip (Load CT → Structures & PTV → Collimator (BEV) → Monte Carlo → Normalize Dose → DVH Evaluation) tracks data readiness and turns each step badge green when its required inputs are present. Complete planning sessions save to a YAML (`.yaml`) study file paired with a compressed NumPy (`.npz`) contour sidecar.

### S2.2 CT, structures and dose visualization

The CT tab displays synchronized axial, coronal, and sagittal multiplanar reconstructions alongside volume and isosurface rendering, with window/level presets and linked crosshairs. DICOM CT series and NIfTI volumes are imported into a canonical LPS orientation and can be re-exported in either format. Structures are imported from DICOM RT Structure Sets or NIfTI label maps or contoured directly in any view with brush and polygon tools. Drawn contours are interpolated across uncontoured slices and exported as NIfTI label maps or DICOM RT Structure Sets. Dose loads from DICOM RT Dose, together with any companion 1σ uncertainty file, and renders as a color wash, isodose lines (default 20, 50, 80, 95, 100, and 110% of prescription), and a translucent 3D volume. Two additional modes display dose rate washes or isodose rate lines (default 10, 40, 100, 200, and 400 Gy/s) based on the pulse timing set in the DVH tab. Interactive 1D line profiles plot dose or dose rate with ±1σ confidence bands alongside Hounsfield units, and a coronal radiochromic film overlay supports planar registration and 2D local or global gamma evaluation.

## S3 BEAM-MODEL VALIDATION

### S3.1 Depth-dependent lateral dose profiles

The solid-water measurements were made in both the UHDR (FLASH) and the conventional dose-rate (CONV) delivery modes of the same linac configuration. The main text reports only the FLASH comparison, which is the mode used for the two end-to-end tests. The CONV films were irradiated at 12 MeV through the same collimator and film stack and were analyzed with the same procedure, and both data sets are shown here for completeness. Note that the beam model described in Sec. S1 was used for both FLASH and CONV comparisons. Figure S1 presents the measured and simulated lateral profiles at nominal water depths of 0, 2, 4, 6, 8, 10, and 12 mm for CONV and FLASH irradiation. The *X* and *Y* profiles were extracted by averaging over a 1-mm-wide strip in the perpendicular direction. For comparisons of field shape, each profile was normalized independently to its mean over the central $-5$ to $+5$ mm interval.

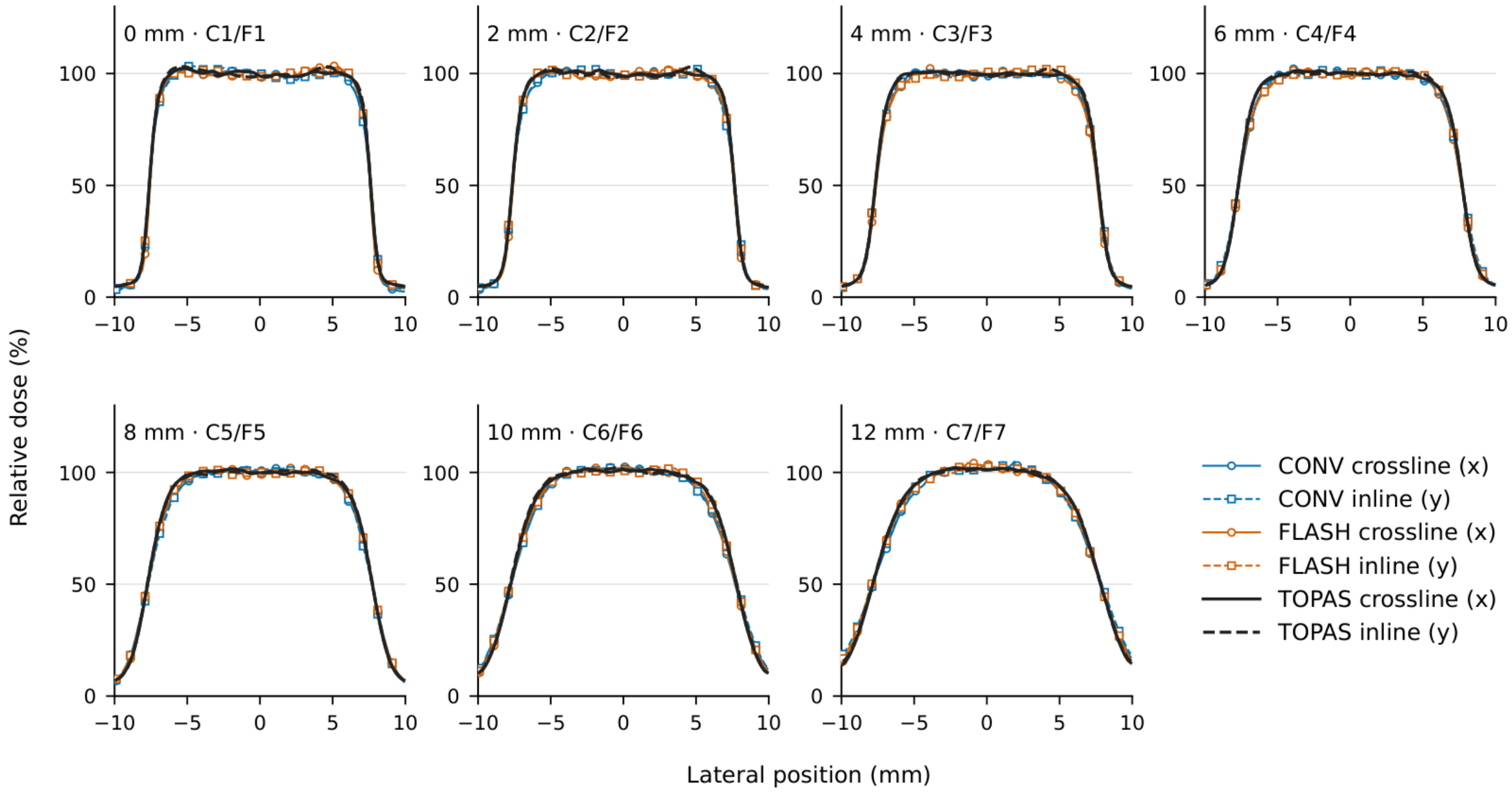

**Figure S1:** Measured EBT-XD film and simulated TOPAS lateral dose profiles at seven nominal water depths for CONV and FLASH irradiation through a $15 \times 15\ \mathrm{mm}^2$ aperture. Crossline (x) and inline (y) profiles are normalized independently to their corresponding central plateau means.

### S3.2 Two-dimensional gamma analysis

The film dose maps were compared with the TOPAS dose distributions using two-dimensional local gamma analysis at 2%/2 mm, 3%/3 mm, and 4%/4 mm.

The measured films served as the reference distributions, and the dose-difference tolerance was calculated relative to the measured dose at each assessed pixel. Pixels with measured doses below 10% of the corresponding film's robust maximum, defined as the 99th percentile of its finite dose values, were excluded. Passing rates were calculated as the percentage of assessed pixels satisfying $\gamma \leq 1$. Figure S2 shows the spatial gamma distributions for all three criteria with the corresponding passing rates.

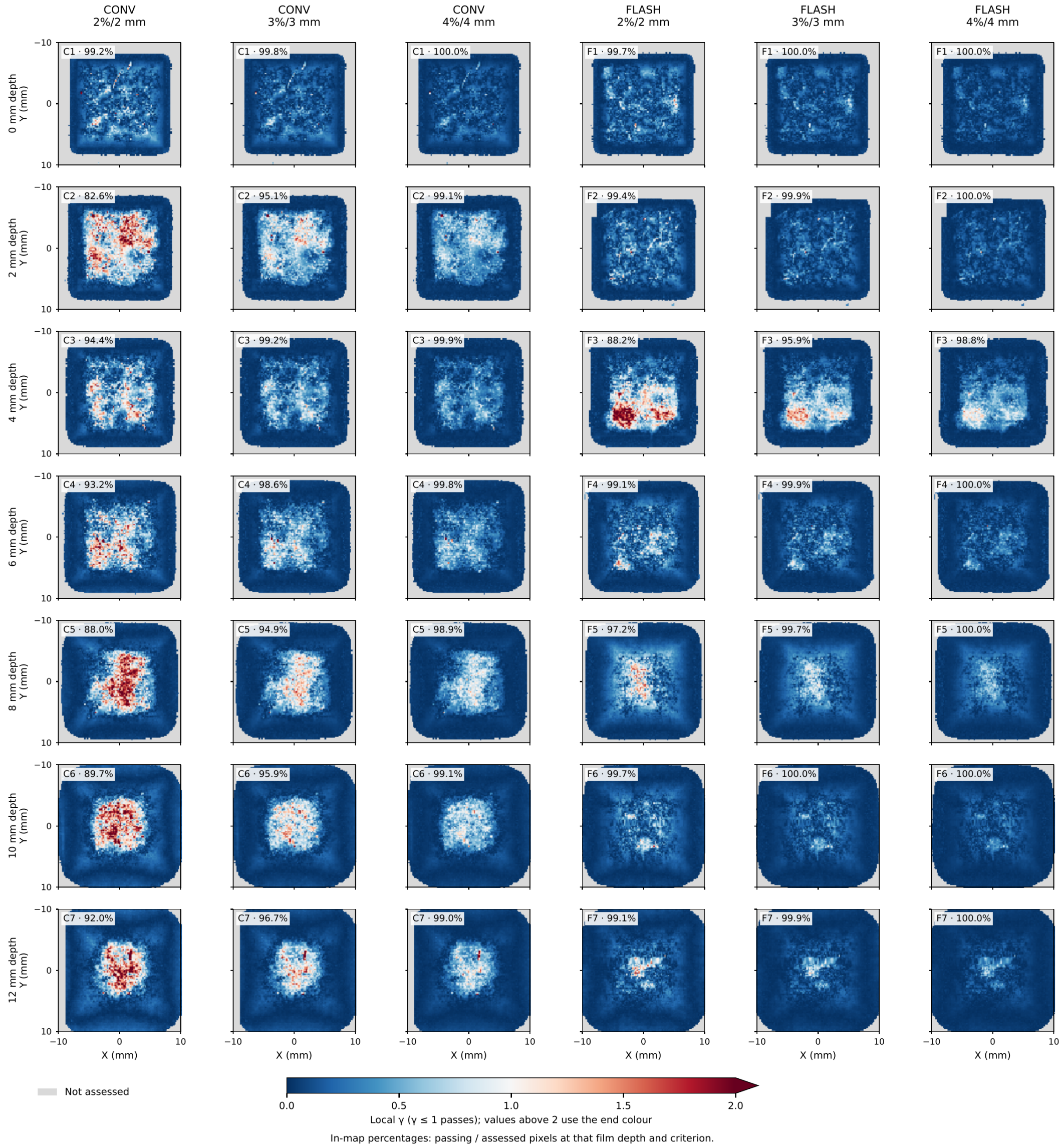

**Figure S2:** Two-dimensional local gamma-index distributions at the seven nominal film depths for (left) CONV and (right) FLASH irradiation. A gamma index $\gamma \leq 1$ denotes agreement within the specified criterion. Excluded regions have insufficient valid film coverage or are below the dose threshold.

## REFERENCES

1. W. Schroeder, K. Martin, and B. Lorensen, *The Visualization Toolkit: An Object-Oriented Approach to 3D Graphics*, 4th ed. (Kitware, Inc., 2006).